\documentclass[12pt]{iopart}
\usepackage{amssymb, amsfonts, amsbsy}
\begin{document}
\def\TODAY{21 December 2010}
\title   
{The causal structure of spacetime is a parameterized Randers geometry}
\author{
Jozef Skakala and 
Matt Visser}
\address{School of Mathematics, Statistics, and Operations Research \\
Victoria University of Wellington, PO Box 600, Wellington, New Zealand}
\ead{jozef.skakala@msor.vuw.ac.nz, 
matt.visser@msor.vuw.ac.nz}
\date{\TODAY; \LaTeX-ed \today}    
\begin{abstract}

There is a by now well-established isomorphism between stationary 4-dimensional spacetimes and 3-dimensional purely spatial Randers geometries --- these Randers geometries being a particular case of the more general class of 3-dimensional Finsler geometries. 
We point out that in stably causal spacetimes, by using the (time-dependent) ADM decomposition, this result can be extended to general non-stationary spacetimes --- the causal structure (conformal structure) of the full spacetime is completely encoded in a parameterized ($t$-dependent) class of Randers spaces, which can then be used to define a Fermat principle, and also to reconstruct the null cones and causal structure. 

\vskip 0.250cm

\noindent
Keywords: Randers norm, Randers metric, causal structure.

\vskip 0.1250cm

\bigskip
\noindent 
\TODAY;  
\LaTeX-ed  \today

\bigskip
\hrule  
\tableofcontents
\bigskip
\hrule
\end{abstract}
\maketitle
\newtheorem{theorem}{Theorem}
\newtheorem{corollary}{Corollary}
\newtheorem{lemma}{Lemma}
\def\d{{\mathrm{d}}}
\def\implies{\Rightarrow}
\newcommand{\scri}{\mathscr{I}}
\newcommand{\sun}{\ensuremath{\odot}}
\def\ep{\epsilon}
\def\k{\mathbf{k}}
\def\x{\mathbf{x}}
\def\v{\mathbf{v}}
\def\s{\mathbf{s}}
\def\e{\mathbf{e}}
\def\t{\mathbf{t}}
\def\n{\mathbf{n}}
\def\u{\mathbf{u}}
\def\w{\mathbf{w}}
\def\eg{{\it e.g.}}
\def\ie{{\it i.e.}}
\def\etc{{\it etc.}}
\def\sign{{\hbox{sign}}}
\def\eof{\Box}
\newenvironment{warning}{{\noindent\bf Warning: }}{\hfill $\eof$\break}
\markboth{The causal structure of spacetime is a parameterized Randers geometry}{}
\section{Introduction}

There is a by now well-established isomorphism between stationary 4-dimensional spacetimes and 3-dimensional purely spatial Randers geometries~\cite{Gibbons:2008, Gibbons:2009, Caponio:2009a, Caponio:2009b} --- these Randers geometries~\cite{Randers} being a particular case of the more general class of 3-dimensional Finsler geometries~\cite{Riemann, Finsler, Buseman, Rund, Chern1, Chern2, Chern3, Chern4, Shen1, Shen2, Antonelli93, Antonelli98, Bejancu}.
This isomorphism is usually established via a projection from the full spacetime to the spatial slices, with the null geodesics of the stationary spacetime projecting down to spatial curves that satisfy Fermat's principle in terms of a specific  anisotropic ``optical metric'' that physicists refer to as a Randers metric~\cite{Randers}, (see also~\cite{Gibbons:2008, Gibbons:2009, Caponio:2009a, Caponio:2009b}).  
We point out that by assuming \emph{stable causality} and using the general (time-dependent) ADM decomposition, this formalism can be extended to general non-stationary spacetimes --- the causal structure (conformal structure) of the full non-stationary spacetime is completely encoded in a parameterized class of Randers spaces, which can then be used to define a Fermat principle, and so reconstruct the null cones. 

\section{From null cones to parameterized Randers norm}

Suppose first that one has a \emph{stably causal} spacetime. Then by definition one has a preferred globally defined time coordinate $t$ (often called a ``cosmic time'') such that $\d t$ is always timelike. If we now perform an ADM decomposition of the metric, using the cosmic time $t$ as one of the coordinates, then in terms of the lapse scalar $N$, shift vector $N^k$, and spatial metric $g_{ij}$, we have
\begin{equation}
g_{ab} =
\left[
\begin{array}{c|c}
- [N^2- g_{kl} N^k N^l ]& g_{jk} N^k  \\
\hline
 g_{ik} N^k  & g_{ij}  \\
\end{array}
\right],
\end{equation}
and
\begin{equation}
g^{ab} =
\left[
\begin{array}{c|c}
- {1/N^2}& N^j/N^2  \\
\hline
 N^i/N^2  & g^{ij} - N^i N^j/N^2\\
 \end{array}
\right],
\end{equation}
where $g^{ij} = [g_{ij}]^{-1}$,  and we can now guarantee $N> 0$ globally. Furthermore, due to the Lorentzian signature of spacetime,  $N>0$ implies that $\det(g_{ij})>0$ globally, so the inverse 3-metric is guaranteed to exist. 
All quantities appearing here can depend implicitly on both space $x^i$, and time $t$. The indices $a$, $b$ lie in $\{0,1,2,3\}$, while the indices $i$, $j$, $k$, $l$ lie in $\{1,2,3\}$.
Stable causality by itself is not quite enough to imply that all the spatial slices have the same topology, but the somewhat stronger condition of \emph{global hyperbolicity} certainly is. For simplicity we shall assume all spatial slices have the same topology. 

\bigskip

\noindent
Now consider a null curve described by
\begin{equation}
\d s^2 = 0 = - N^2 \d t^2 + g_{ij} (\d x^i + N^i \d t) (\d x^j + N^j \d t).
\end{equation}
That is
\begin{equation}
- [N^2 - g_{kl} N^k N^l] \d t^2 +2 N_k \d x^k \d t + g_{ij} \d x^i \d x^j = 0.
\end{equation}
This can be viewed as a simple quadratic for $\d t$, yielding
\begin{equation}
\d t =  { N_k \d x^k \pm \sqrt{ ( N_k \d x^k)^2 + ( g_{ij} \d x^i \d x^j)(N^2 - g_{kl} N^k N^l) }\over [N^2 - g_{kl} N^k N^l] }
\end{equation}
More explicitly
\begin{equation}
\fl
\d t =  { N_k \d x^k \over [N^2 - g_{kl} N^k N^l] } 
\pm \sqrt{  \left[ {g_{ij} \over N^2 - g_{kl} N^k N^l}  + {N_i N_j \over [N^2 - g_{kl} N^k N^l]^2} \right] \d x^i \d x^j  }.
\end{equation}
Define a vector
\begin{equation}
V_k = { N_k \over [N^2 - g_{kl} N^k N^l] } ,
\end{equation}
and a matrix
\begin{equation}
h_{ij} =  {g_{ij} \over N^2 - g_{kl} N^k N^l}  + {N_i N_j \over [N^2 - g_{kl} N^k N^l]^2} =  {g_{ij} \over N^2 - g_{kl} N^k N^l} + V_i V_j.
\end{equation}
As long as 
\begin{equation}
N^2 - g_{kl} N^k N^l > 0, \qquad \hbox{that is} \qquad  g_{kl} N^k N^l  < N^2,
\end{equation}
then the $3\times3$ matrix $h_{ij}$ will be positive definite (and so invertible). 

From an ``analogue spacetime'' perspective~\cite{Unruh, unexpected, ergo, LRR} this corresponds to a situation where the flow velocity is less than the propagation speed of whatever signal one is considering. It is thus equivalent, from  an ``analogue spacetime'' perspective,  to the condition that there be no ``ergoregion'' in this particular coordinate system~\cite{ergo, LRR}.  

In a more abstract purely general relativistic setting, this is a condition that the shift vector be subliminal, so that the coordinate system is not being stretched too much from one time-slice to the next.
Furthermore, in this situation we are guaranteed that
\begin{equation}
\label{E:invertible}
  h_{ij} \, \d x^i \d x^j > (N_i \, \d x^i)^2;  \qquad (\d x^i \neq 0).
\end{equation}

\bigskip

Now define the quantity
\begin{equation}
R_\pm(t,x^m,\d x^m) =  \sqrt{  h_{ij}(t, x^m) \,\d x^i \d x^j }   \pm V_i(t,x^m) \,\d x^i \geq 0.
\end{equation}
More abstractly let $v^m$ be the components of a spatial 3-vector residing in the spatial tangent space at $(t, x^m)$ and define
\begin{equation}
R_\pm(t,x^m, v^m) =  \sqrt{  h_{ij}(t, x^m) \,v^i v^j }   \pm V_i(t,x^m) \,v^i \geq 0.
\end{equation}
Then by construction the quantity $R(t,x^m, v^m) $ is linear in the components $v^i$, that is,
\begin{equation}
R_\pm(t,x^m, \lambda \; v^m) =  \lambda \; R(t,x^m, v^m),
\end{equation}
and furthermore  $R_\pm(t,x^m, v^m) =0$ if and only if $v^m=0$. These are the defining characteristics of a Finsler norm [defined on the purely spatial tangent space at $(t,\vec x)$].  If $V_i(t,\vec x)=0$ (which happens if and only if the shift vector $N^i(t,\vec x)$ is zero) then this Finsler norm in fact reduces to a purely spatial Riemannian norm. For $V_i(t,\vec x)\neq 0$ this particular Finsler norm is an example of a special subset of Finsler norms referred to as a Randers norm~\cite{Randers}. These Randers norms were first developed in a physics context, as an attempt to generalize Einstein gravity~\cite{Randers}. 
In contrast, the intent here is rather different --- within the context of \emph{standard} Lorentzian spacetime Randers norms on the \emph{spatial slices} are being used to encode some of the information content of the spacetime metric. 

That some information is unavoidably lost in going from the spacetime metric to the Randers norm is physically clear from the fact that in constructing these particular Randers norms we have only used the null curves. Mathematically this is clear from the fact that the Randers norms are conformal invariants of the underlying spacetime metric. The conformal transformation $g_{ab} \to \Omega^2 \; g_{ab}$ leaves both $h_{ij}$ and $V_i$ invariant.

\section{Fermat's principle}

Physically the interpretation of the Randers norm is clear. We have
\begin{equation}
\d t =  \pm R_\pm(t,x^m,\d x^m),
\end{equation}
so the Randers norm is telling us how much time (cosmic time $t$) it takes for a light signal to move a coordinate distance $\d x^i$.
If we consider a curve $\gamma$ in space parameterized by $\sigma$, so that the coordinate representation of  $\gamma$ is $x^i(\sigma)$, then we can lift this curve to spacetime by considering $x^a(\sigma) = (t(\sigma),x^i(\sigma))$. Then in order for the spacetime curve to be a null curve we must have
\begin{equation}
{\d t\over\d \sigma} = \pm R_\pm\left(t(\sigma),x^m(\sigma),{\d x^m(\sigma)\over\d \sigma}\right).
\end{equation}
This is a first-order ODE for $t(\sigma)$ that completely (if implicitly) determines $t(\sigma)$ in terms of the spatial curve $\gamma$. 
Then one can define the total time taken to traverse the spacetime curve as
\begin{equation}
T(\gamma) = \int_\gamma \d t =   \pm \int  R_\pm\left(t(\sigma),x^m(\sigma),{\d x^m(\sigma)\over\d \sigma}\right) \d \sigma.
\end{equation}
Note that $T(\gamma)$ (because of the homogeneity property of the Randers norm) is independent of any re-parameterization $\sigma \to f(\sigma)$.  The $\pm$ simply has to do with the direction in which one traverses the curve.  Extremizing $T(\gamma)$ is just Fermat's principle.  For explicit discussion of the stationary case see for example~\cite{Gibbons:2008}, or earlier expositions in~\cite{unexpected, ergo, LRR} and~\cite{Fluids, Classical}. Checking that the discussion continues to hold in non-steady situations is straightforward.  We note that by construction $T(\gamma)$ can equally well be written in terms of $x^a(\sigma)$ as:
\begin{equation}
T(\gamma) = \int_\gamma {\d x^a \over\d\sigma} \; \nabla_a t   \; \d \sigma 
\qquad \hbox{subject to} \qquad 
g_{ab}  {\d x^a \over\d\sigma}  {\d x^b \over\d\sigma} = 0.
\end{equation}
Introducing the Lagrange multiplier $\Lambda(\sigma)$ one has an \emph{equivalent} variational principle 
\begin{equation}
\tilde T(\gamma) = \int_\gamma \left[ {\d x^a \over\d\sigma} \; \nabla_a t   + \Lambda(\sigma) g_{ab}  {\d x^a \over\d\sigma}  {\d x^b \over\d\sigma} \right]\; \d \sigma. 
\end{equation}
But for this equivalent variational principle
\begin{equation}
\tilde T(\gamma) =  t(\sigma_f) - t(\sigma_i) + \int_\gamma \Lambda(\sigma) \; g_{ab}  {\d x^a \over\d\sigma}  {\d x^b \over\d\sigma} \; \d \sigma. 
\end{equation}
This last variational principle will now (as required) yield the standard null geodesic equations.

\section{From parameterized Randers norm to metric null cones}

Can we now reconstruct the spacetime metric (or at least its conformal class) from the Randers norm?  By making measurements of $\d t$ for light rays moving in different directions one can in principle extract both $h_{ij}$ and $V_i$.  To now (partially) reconstruct the spacetime metric note that
\begin{equation}
g_{ij} = [N^2 - g_{kl} N^k N^l] \left\{ h_{ij} - V_i V_j \right\},
\end{equation}
and
\begin{equation}
N_{i} = [N^2 - g_{kl} N^k N^l] \left\{ V_i \right\}.
\end{equation}
Consequently there exists a scalar $\Omega(t,\vec x)$ such that
\begin{equation}
g_{ab} = \Omega^2(t,\vec x)
\left[
\begin{array}{c|c}
- 1 & V_j  \\
\hline
V_i  & h_{ij} - V_i V_j \\
\end{array}
\right].
\end{equation}
Furthermore, as long as condition (\ref{E:invertible}) is satisfied, then $h_{ij} - V_i V_j$ is positive definite, and so invertible.  Subject to this condition it is useful to define $\hat g_{ij} = h_{ij} - V_i V_j$, so that
\begin{equation}
g_{ab} = \Omega^2(t,\vec x)
\left[
\begin{array}{c|c}
- 1 & V_j  \\
\hline
V_i  & \hat g_{ij} \\
\end{array}
\right].
\end{equation}
But now let us define $\hat g^{ij} = [\hat g_{ij}]^{-1}$ to be the inverse matrix to $\hat g_{ij}$, and further define
\begin{equation}
\hat N^2 = 1 + \hat g^{ij} V_i V_j \geq 1.
\end{equation}
This allows us to reconstruct the spacetime metric $g_{ab}$ in ``conformally ADM'' form
\begin{equation}
g_{ab} = \Omega^2(t,\vec x)
\left[
\begin{array}{c|c}
- [\hat N^2 -   \hat g^{ij} V_i V_j] & V_j  \\
\hline
V_i  & \hat g_{ij} \\
\end{array}
\right].
\end{equation}
Finally, by defining
\begin{equation}
\hat V^i = \hat g^{ik} V_k,
\end{equation}
where the $\hat V^i$ notation is needed to avoid any possibility of ambiguity as to which particular 3-metric is being used to raise the index, we see that the inverse spacetime metric is
\begin{equation}
g^{ab} = \Omega^{-2}(t,\vec x)
\left[
\begin{array}{c|c}
- 1/ \hat N^{2}&  \hat V^j/\hat N^{2} \\
\hline
\vphantom{\bigg|} 
\hat V^i /\hat N^{2} & \hat g^{ij} - \hat V^i \hat V^j /\hat N^{2}\\
\end{array}
\right],
\end{equation}
thereby completing the reconstruction (up to a conformal factor) of the spacetime metric in terms of the Randers norm (or more specifically the  3-dimensional quantities $h_{ij}$ and $V_i$).
Note in particular that this implies that the (3+1) Weyl tensor is in principle completely calculable in terms the  3-dimensional purely spatial quantities $h_{ij}$ and $V_i$. Explicit computation is, like Finsler space computations generally, likely to be tedious.

\section{From Randers norm to Randers metric}

The step from Randers norm to Randers metric is straightforward but tedious.  Since a Randers geometry is just a special case of a Finsler geometry one defines a direction-dependent metric by 
\begin{equation}
[g_R]_{ij}(t,x^k,v^k) =  {1\over2}  {\partial^2 [R^2_\pm]\over\partial v^i \partial v^j}.
\end{equation}
That is
\begin{equation}
[g_R]_{ij}(t,x^k,v^k) =  {1\over2}  {\partial^2 [h_{kl} v^k v^l + (V_k v^k)^2 \pm 2 \sqrt{h_{kl} v^k v^l } (V_k v^k)]\over\partial v^i \partial v^j}.
\end{equation}
Thus
\begin{equation}
[g_R]_{ij}(t,x^k,v^k) =   h_{ij} + V_i V_j  \pm {\partial^2[ \sqrt{h_{kl} v^k v^l } (V_k v^k)]\over\partial v^i \partial v^j}.
\end{equation}
A brief computation now yields
\begin{eqnarray}
\fl
[g_R]_{ij}(t,x^k,v^k) &=&   h_{ij} + V_i V_j  
\nonumber\\
\fl &\pm& \left\{  
h_{ij} {V_k v^k\over \sqrt{h_{kl} v^k v^l }} 
-{(h_{ik} v^k)\,(h_{jl} v^l)\, (V_k v^k) \over ({h_{kl} v^k v^l })^{3/2}}   
+{(h_{ik} v^k) V_j + V_i (h_{jk} v^k)\over  \sqrt{h_{kl} v^k v^l }}
\right\},
\end{eqnarray}
which has the usual Finsler interpretation of a ``direction dependent 3-metric''.
Note that (as required for any Finsler metric) this is explicitly zero-order homogeneous in the tangent vector $v^k$. That is 
\begin{equation}
[g_R]_{ij}(t,x^k, \lambda v^k) = [g_R]_{ij}(t,x^k,v^k).
\end{equation}
Furthermore, note that because the spacetime underlying the Randers norm is mono-metric, with a single null cone, the issues and problems arising in~\cite{lnp, Jozef1,  Jozef2, Jozef3} are not a issue. The only place where the Randers metric fails to be smooth is at $v^k=0$, so we have the usual restriction that the Randers metric is defined on the so-called ``slit tangent bundle'' (the tangent bundle excluding the zero vector). 

\section{Relation to $\alpha$-$\beta$ norms}

When comparing to the mathematical literature, it is useful to realise that Randers norms are a special case of the so-called $\alpha$-$\beta$ norms, which are themselves still a special case of general Finsler norms.  It is standard to define
\begin{equation}
\alpha =  \sqrt{h_{ij} v^i v^j};   \qquad \beta = V_k v^k.
\end{equation}
Any function $f(\alpha, \beta)$ that satisfies the homogeneity property
\begin{equation}
f(\lambda \alpha, \lambda\beta) = \lambda f(\alpha, \beta),
\end{equation}
then defines an  $\alpha$-$\beta$ norm 
\begin{equation}
F(x,v) = f(\alpha,\beta).
\end{equation}
The Randers norm corresponds to $f(\alpha, \beta)= \alpha+\beta$, while $f(\alpha, \beta)=\alpha$ in isolation corresponds to a Riemannian norm, and the combination $f(\alpha, \beta)= \alpha + \beta^2/\alpha$ is yet another example. The so-called Kropina norm corresponds to $f(\alpha, \beta)=\alpha^2/\beta$ and the Matsumoto norm to $f(\alpha, \beta)=\alpha^2/(\alpha-\beta) = \alpha^2 (\alpha+\beta)/(\alpha^2-\beta^2)$. Among this plethora of possibilities it is the Riemannian norm and the Randers norm  that appear to be the most useful for physics applications.

\section{Discussion and Conclusions}

The central result of this article is that the notion of 3-dimensional Randers space is useful not only for stationary spacetimes, but that under mild conditions (such as stable causality and a mild constraint on the shift vector) time-dependent Randers norms can usefully  be employed to characterize the causal structure of general spacetimes.  We have given an explicit construction of the Randers norm in terms of the underlying spacetime metric, and have explicitly shown how to reconstruct (up to an intrinsically undetermined conformal factor) the spacetime metric from the spatial Randers norms. The construction serves to directly connect more abstract parts of the mathematics literature with issues of direct relevance to physics.  In particular we hope that this construction will provide a new point of view regarding causal structures in Lorentzian geometries. 

\ack

This research was supported by the Marsden Fund administered by the
Royal Society of New Zealand. 
JS was also supported by a Victoria University of Wellington postgraduate scholarship.

\section*{References}
\addcontentsline{toc}{section}{References}


\end{document}